\documentclass[sigconf]{acmart}

\usepackage{blindtext, graphicx}
\usepackage{subfigure}
\usepackage{amsmath}
\usepackage{amssymb}

\AtBeginDocument{%
  \providecommand\BibTeX{{%
    \normalfont B\kern-0.5em{\scshape i\kern-0.25em b}\kern-0.8em\TeX}}}
\setcopyright{acmcopyright}
\copyrightyear{2019}
\acmYear{2019}
\acmDOI{10.1145/1122445.1122456}
\acmConference[ICONS '19]{ICONS '19: ACM International Conference on Neuromorphic Systems}{July 23--25, 2019}{Knoxville, Tn}
\acmBooktitle{ICONS '19: ACM International Conference on Neuromorphic Systems, Knoxville, Tn}
\acmPrice{15.00}
\acmISBN{978-1-4503-9999-9/18/06}
\begin{document}
\title{A Low-Power Domino Logic Architecture for Memristor-Based Neuromorphic Computing}
\author{Cory Merkel}
\email{cemeec@rit.edu}
\author{Animesh Nikam}
\email{an8098@rit.edu}
\affiliation{%
  \institution{Brain Lab\\Rochester Institute of Technology}
  \streetaddress{1 Lomb Memorial Drive}
  \city{Rochester}
  \state{New York}
  \postcode{14624}
}
\renewcommand{\shortauthors}{Merkel and Nikam}

\begin{abstract}
We propose a domino logic architecture for memristor-based neuromorphic computing.  The design uses the delay of memristor RC circuits to represent synaptic computations and a simple binary neuron activation function.  Synchronization schemes are proposed for communicating information between neural network layers, and a simple linear power model is developed to estimate the design's energy efficiency for a particular network size.  Results indicate that the proposed architecture can achieve 0.61 fJ per classification per component (neurons and synapses) and outperforms other designs in terms of energy per \% accuracy.
\end{abstract}




\maketitle

\vspace{-2mm}
\section{Introduction}

Custom neuromorphic hardware platforms are gaining popularity for the acceleration of neural network algorithms, owing to their ability to perform complex tasks that are analogous of the physical processes underlying biological nervous systems \cite{neuromorphic}. A key feature of these systems is that they overcome the limitations caused by the von Neumann bottleneck by collocating computation and memory \cite{brain}. While modern digital complementary-metal-oxide-semiconductor (CMOS) technology is used to replicate the behavior of the neurons, the absence of a device that can efficiently perform synaptic operations stunted progress for several years. However, recent advancements in nanoscale materials and realization of devices such as memristors have opened possibilities for developing compact memory device arrays that are potentially transformative for the design of ultra energy-efficient neuromorphic systems.  

Previous work has studied several aspects of memristor-based neuromorphic systems, including device properties, reliability, crossbar implementation, on-chip training, quantization, and much more \cite{sung2018perspective,schuman2017survey}.  One of the most power-efficient design approaches is combining memristor synapses with an integrate-and-fire (IF) neuron design.  The energy efficiency of the IF neuron comes from i.) all-or-nothing representation of information and ii.) little-to-no short-circuit current between the neuron's input and the synapses driving it (since they are just driving the membrane capacitor).  In this work, we explore a similar idea applied to networks of binary neurons inspired by domino logic.  Domino logic, a type of dynamic logic, separates a circuit into pre-charge and evaluation phases to avoid short circuit current and reduce power consumption.  Here, we propose a domino logic style neuron that uses memristor-based RC delays for evaluation and offers good power efficiency.  The building blocks of the proposed design are outline in Section \ref{sec:building}.  Then, Section \ref{sec:multi} discusses scaling the design up to a multi-layer neural network.  In Sections \ref{sec:power} and \ref{sec:quant}, we detail the power consumption model and quantization approach.  Section \ref{sec:results} discusses results on the MNIST dataset and concludes this work.

\section{Circuit Building Blocks}
\label{sec:building}

\begin{figure*}[!t]
\centering
\subfigure[]{
\includegraphics[width=0.35\columnwidth]{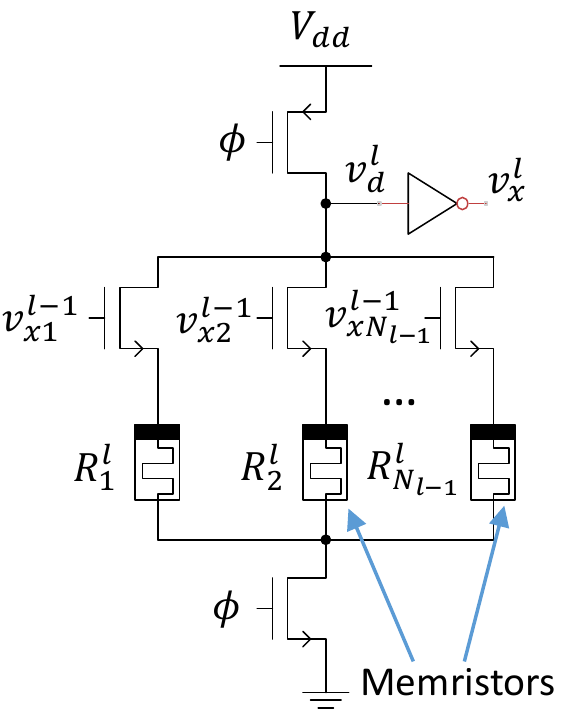}
\label{fig:domino_neuron}
}
\hspace{-2mm}
\subfigure[]{
\raisebox{1.5mm}{
\includegraphics[width=0.45\columnwidth]{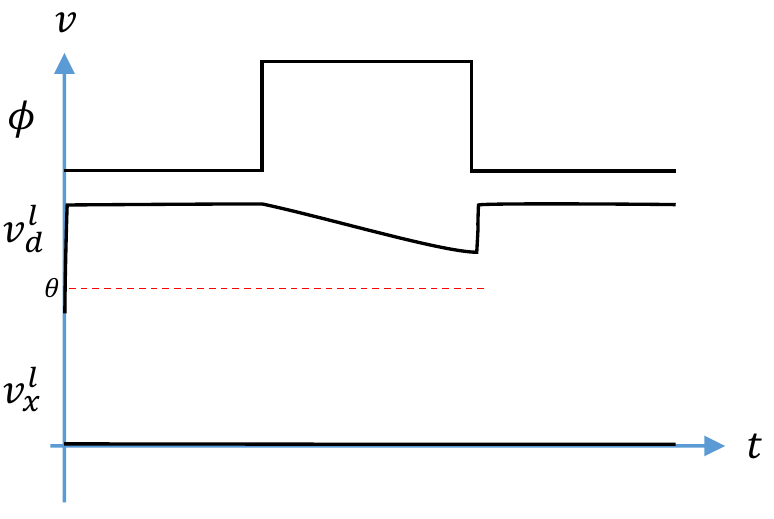}}
\label{fig:dominoneuron0}
}
\hspace{-2mm}
\subfigure[]{
\raisebox{1.5mm}{
\includegraphics[width=0.45\columnwidth]{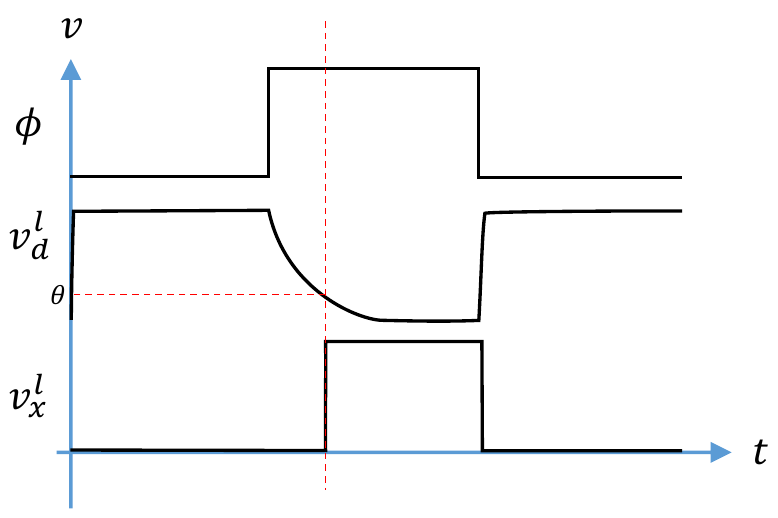}}
\label{fig:dominoneuron1}
}
\hspace{-2mm}
\subfigure[]{
\raisebox{1.5mm}{
\includegraphics[width=0.45\columnwidth]{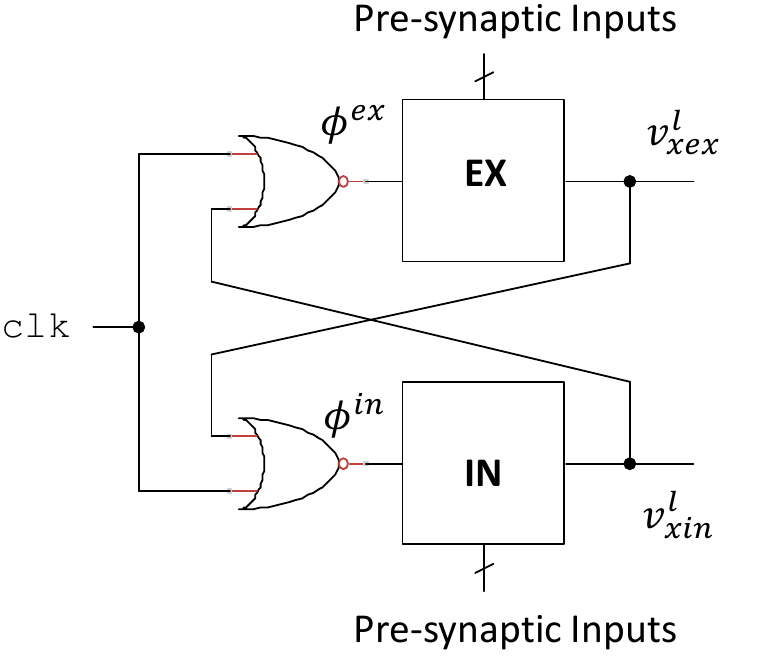}}
\label{fig:inhibneuron}
}
\vspace{-4mm}
\caption{(a) Domino logic-style neuron circuit schematic.  (b)  The neuron's dynamic node does not cross the inverter threshold before the end of the evaluation period, indicating a '0'.  (c)  The neuron's dynamic node reaches the inverter threshold within the evaluation period, indicating a '1'.  (d) Combining two domino circuits to create synapses that can be positive or negative.  
}
\vspace{-4mm}
\end{figure*}

The core building block of our design is shown in Figure \ref{fig:domino_neuron}.  When the clock signal $\phi$ is low, the dynamic node (input to the inverter) is precharged to $V_{dd}$.  Then, during the evaluation phase, $\phi$ is high, and the dynamic node discharges at a rate dependent on the pull-down network's RC time constant.  Once the dynamic node falls below the inverter threshold, the output will go high. 

During the evaluation phase, the voltage on the dynamic node evolves as
\begin{equation}
v_{di}^{l}(t)=V_{dd}\mathrm{exp}\left(-\int\limits_{0}^{t}\frac{G_{i}^{l}(\xi)}{C}\mathrm{d}\xi\right)
\label{eqn:vd}
\end{equation}
where $G_{i}^{l}$ is the equivalent pulldown conductance.  A memristor only contributes to the pull-down conductance when its select transistor is on.  Assuming that memristor conductance values are constant during the evaluation phase and input voltages are digital, i.e. $v_{xj}^{l-1}\in \left\lbrace0,V_{dd}\right\rbrace$, then $G_{i}^{l}$ is a piecewise constant function written as:
\begin{equation}
G_{i}^{l}(t)=\frac{1}{V_{dd}}\sum\limits_{j=1}^{N_{l-1}}v_{xj}^{l-1}(t)G_{ij}^{l}
\label{eqn:G}
\end{equation}

In this work, the inputs to a neuron are constant during each evaluation period.  In addition, each neuron's output will be considered as '1' if it switches from '0' to '1' at any point during the evaluation period.  Otherwise, it is '0'.  This is illustrated in Figures \ref{fig:dominoneuron0} and \ref{fig:dominoneuron1}.  In Figure \ref{fig:dominoneuron0}, the neuron's dynamic node does not discharge to the inverter threshold before the end of the evaluation period, so the output is '0'.  In contrast, the dynamic node in Figure \ref{fig:dominoneuron1} discharges quickly, well before the end of the evaluation period, so its output is '1'.



In order to get an inhibitory effect on the post-synaptic neuron, we introduce a second domino circuit, as shown in Figure \ref{fig:inhibneuron}, where the two boxes represent the circuit in Figure \ref{fig:domino_neuron}.  The top domino circuit is an excitatory neuron, where large memristor conductances will tend to cause the excitatory output to be '1' and the inhibitory output to be '0'.  The bottom domino circuit is an inhibitory neuron, where large memristor conductances will tend to cause the inhibitory output to be '1' and the excitatory output to be '0'.  The circuit uses a built-in arbiter (cross-coupled NOR gates) to decide which neuron reached its inverter threshold first and then re-charge the dynamic node of the other neuron so its output will be '0'.  One design issue that should be considered in future work is the detection and cancellation of metastability in the feedback loop, as it could cause unstable behavior and larger power consumption.  On the other hand, it may be a useful tool for implementing stochastic neuorn behavior.  However, this is outside the scope of the present work.

Now, we can define a linear mapping between a weight between -1 and 1 and the two conductance values associated with it:
\begin{equation}
G_{i,j_{ex}}^{l}=\mathrm{max}\left(G_{min},w_{i,j}^{l}G_{max}+\left(1-w_{i,j}^{l}\right)G_{min}\right)
\end{equation}
\begin{equation}
G_{i,j_{in}}^{l}=\mathrm{max}\left(G_{min},-w_{i,j}^{l}G_{max}+\left(1+w_{i,j}^{l}\right)G_{min}\right)
\end{equation}
The above two equations set the inhibitory conductance to $G_{min}$ when the weight is positive and the excitatory conductance to $G_{min}$ when the weight is negative.  Then, the excitatory conductance will range from $G_{min}$ for a weight of 0 to $G_{max}$ for a weight of +1.  The inhibitory conductance will range from $G_{min}$ for a weight of 0 to $G_{max}$ for a weight of -1.  Note that $G_{min}$ and $G_{max}$ are the minimum and maximum conductance values of the memristors and vary considerably based on the type of device (i.e. material properties, fabrication process, etc.) \cite{burr2017neuromorphic}.  In this work, we have chosen values of $G_{min}=1/10^{6}\Omega$ and $G_{max}=1/10^{5}\Omega$, however our design will work for other conductance ranges.

\vspace{-3mm}
\section{Multi-Layer Networks}
\label{sec:multi}

\subsection{Crossbar Implementation}

\begin{figure}
\centering
\includegraphics[width=\columnwidth]{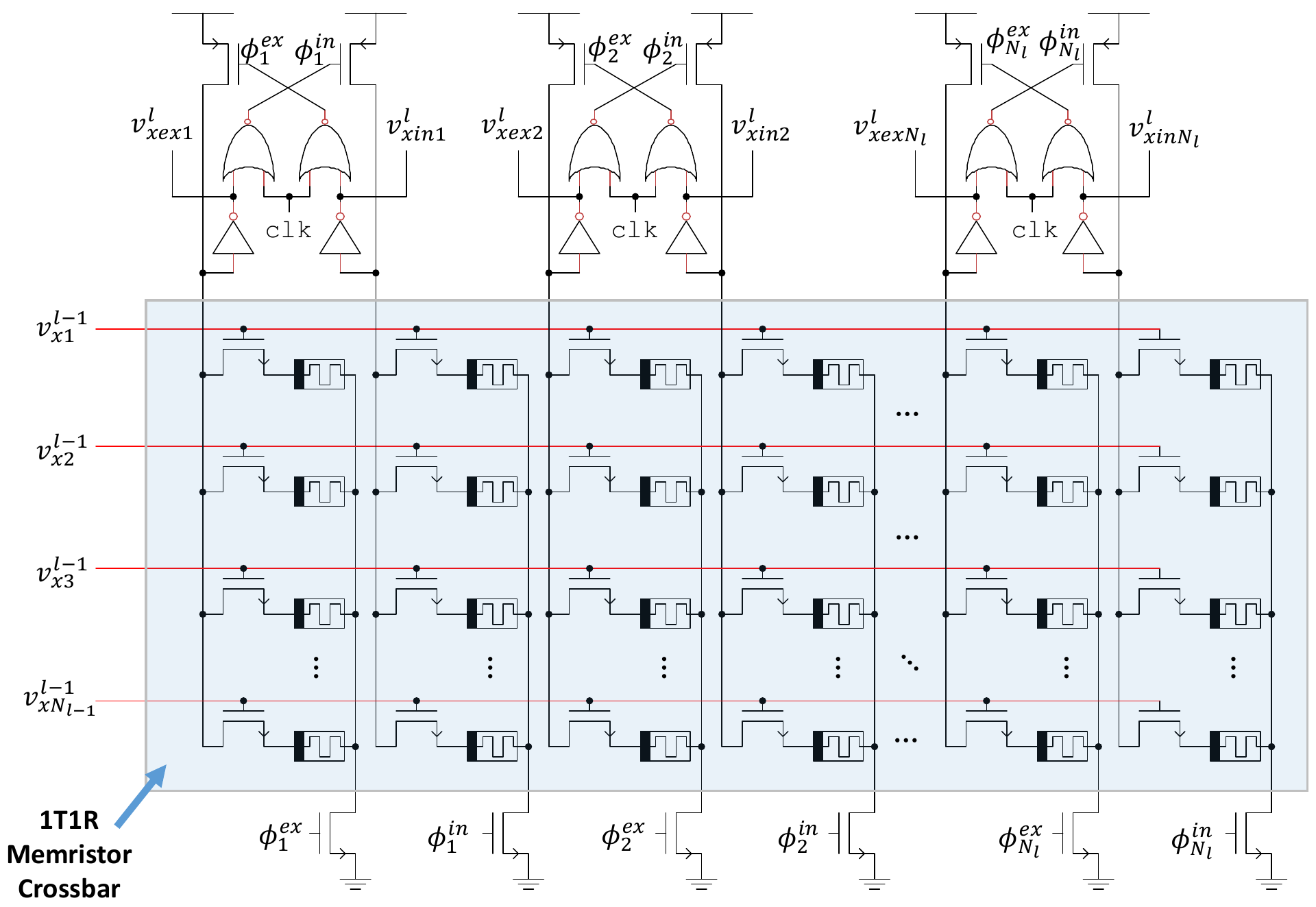}
\caption{Implementation of the synaptic weight matrix between two neural network layers using the proposed neuron design and a 1T1R memristor crossbar.}
\label{fig:xbar}
\vspace{-4mm}
\end{figure}

For multilayer neural networks, we propose a 1T1R memristor crossbar, as shown in Figure \ref{fig:xbar}.  Here, the word lines are connected to the pre-synaptic neuron outputs from the previous layer.  The two terminals of each 1T1R synapse are connected to the crossbar columns.  Each neuron uses two crossbar columns to implement excitatory and inhibitory synapses.  Footer transistors are used to eliminate short circuit power consumption during pre-charge.  Note that secondary pre-charge transistors may be needed to avoid charge sharing between each domino circuit's dynamic node and the drain of the footer transistors.  For fully-connected neural networks, the simplest design would employ one $N_{l-1}\times N_{l}$ crossbar for each layer $l$.  However, more advanced methods will likely be needed for sharing crossbars across layers and efficiently mapping sparse connectivity networks to dense crossbar structures.

\subsection{Synchronization Across Layers}

\begin{figure}
\centering
\includegraphics[width=0.7\columnwidth]{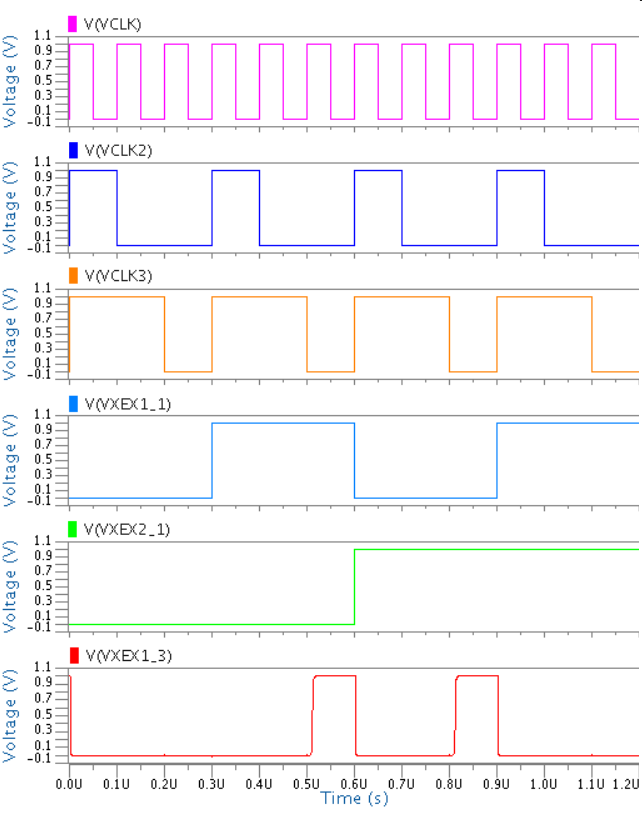}
\caption{MLP simulation of XOR with sequential evaluation across layers.}
\label{fig:xor2}
\vspace{-4mm}
\end{figure}

The proposed design is based on the timing of RC delays in each domino circuit.  Since each neuron's output is binary, it is important that the domino circuits do not perform an evaluation until all of their inputs are ready (i.e. the evaluation period of the inputs has completed).  For this reason, it is critical to have some form of synchronization across layers.  We propose three different methods.  The first method uses non-overlapping clocks  with varying duty cycles for each network layer in the following manner:  First, all of the clocks are '1' to pre-charge all of the domino circuits.  Next, the clock for the first layer becomes '0' for evaluation of the network inputs.  After enough time has passed for evaluation of the first layer (this will depend on the size of the network, weight values, etc.), the clock for the second layer will become '0', and so on until the clock for the final layer becomes '0'.  Then, the process starts over.  The advantage of this approach is that no circuitry has to be added to the neuron circuits.  This disadvantage is that each layer has to wait for all of the previous layers to finish before it can perform any computation.  The second synchronization method is to add flipflops to the output of each neuron.  This way the entire network can be pipelined across layers and each neuron can perform computations on every clock cycle.  Of course, the disadvantage of this approach is that it adds overhead to the neuron design.  A final method is to use asynchronous handshaking across layers.  In this case, a global reset signal would be asserted every time a new input arrives to the network, causing all domino circuits to be pre-charged.  Then, an OR gate would be connected to each neuron's excitatory and inhibitory outputs.  Once the OR gate's output becomes '1', we know that the neuron has finished evaluation.  When all such signals for a whole layer become '1' (which could be detected with an AND tree), that layer has finished evaluating, and the next layer can continue evaluation.  The main advantage of this approach is that a global clock is not needed, which may significantly reduce power consumption.  In this work, we performed simulations using the first method. Figure \ref{fig:xor2} shows the simulation results for a 2-input network with 2 hidden layer neurons and 1 output.  The network was trained to perform the XOR function of its inputs.  The top subplot shows the clock signal, while the second two subplots show the clocks distributed to layers 2 and 3, repsectively.  During the first clock cycle, all of the neurons in the network pre-charge.  During the second clock cycle, the second layer clock goes low, and then the third layer clock goes low during the third clock cycle.  Therfore, the output of the network has a valid result after three cycles from the time that the input changes.


\vspace{-4mm}
\section{Power Consumption}
\label{sec:power}

\begin{figure}
\centering
\includegraphics[width=0.6\columnwidth]{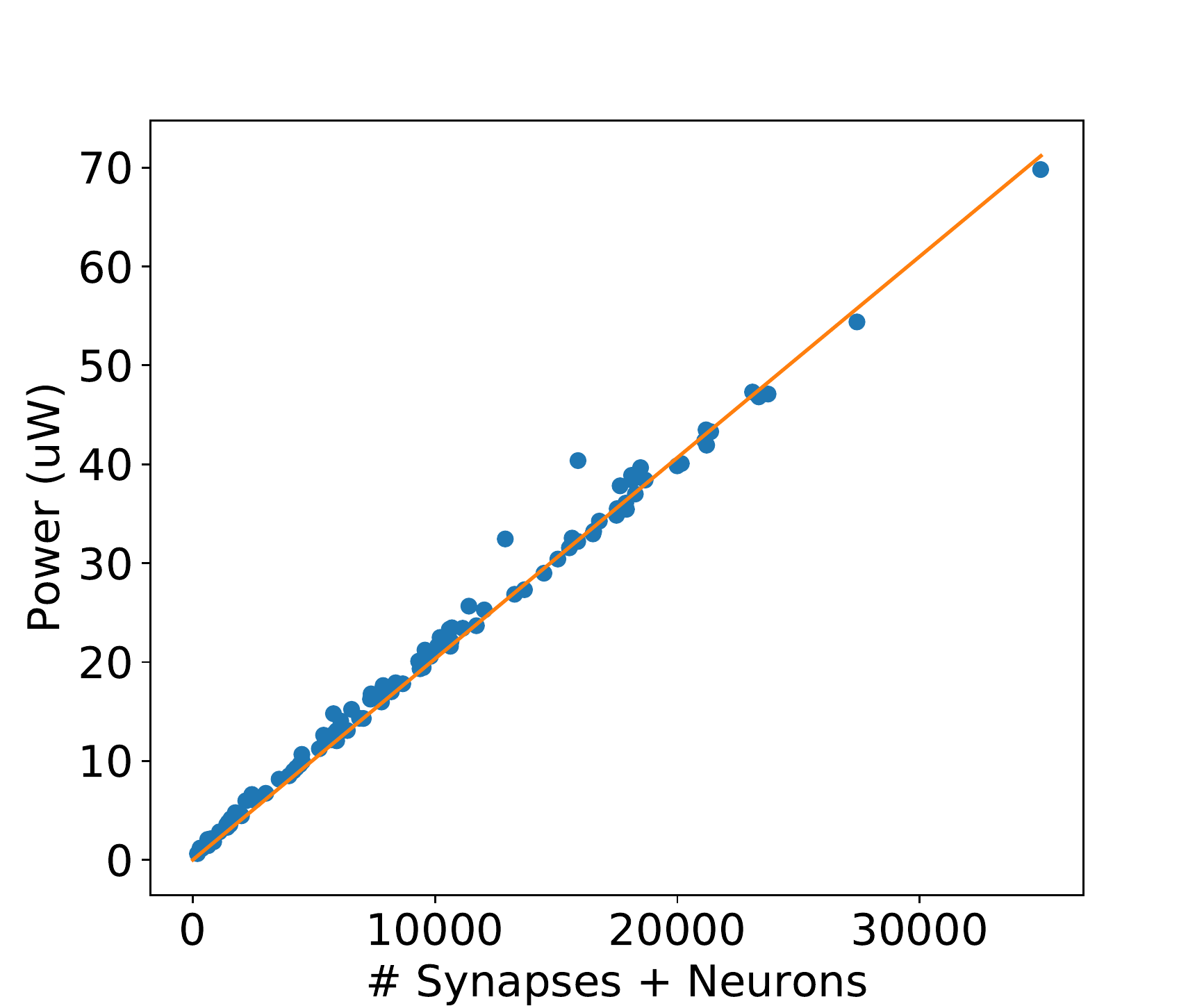}
\caption{Power consumption vs. network size (synapses plus neurons) for the proposed design.}
\label{fig:power_vs_comps}
\vspace{-4mm}
\end{figure}

The power consumption of the proposed design was modeled by assuming that most of the power is consumed when a neuron pre-charges.  The justification for this is that, especially for neurons with high fan-in the switching capacitance of the neuron's dynamic node will be much larger than the capacitance at other nodes in the circuit.  Therefore, the power can be formulated as

\begin{equation}
P\approx3\times(1+\beta)\sum\limits_{L=2}^{N_{L}}\alpha C_{L}V_{dd}^{2}f
\label{eqn:power}
\end{equation}
where $\beta$ is a fitting parameter that comes from the extra power associated with the inverter, arbiter, etc., $\alpha$ is the switching activity factor, and $C_{L}$ is the total switching capacitance of the layer.  The factor of 3 comes from the fact that each synapse will have approximately 3 units of capacitance associated with it from the access transistor's source, drain, and the memristor itself.  Note that a unit of capacitance is calculated as $C_{unit}=A_{fet}\times3.9\times\epsilon_{0}/t_{ox}$, where $A_{fet}$ is the transistor channel area, $\epsilon_{0}$ is the permittivity of free space, and $t_{ox}$ is the transistor gate oxide thickness.  For an $L$ layer network with the chosen synchronization scheme, $\alpha=1/L$, since the neuron circuit only pre-charge once every $L$ clock cycles.  In addition, the value of $C_{L}$ is $C_{unit}$ times the sum of the number of synapses and neurons of each layer (both excitatory and inhibitory).  We have empirically found $\beta\approx 1.15$.  In Figure \ref{fig:power_vs_comps}, we show the power consumption for 100 randomly-sized 3-layer networks vs. the number of synapses and neurons in the network.  For each network, both the inputs and weights were generated randomly.  Furthermore, the network used a clock frequency of 10 MHz.  The results are based on a 130 nm bulk CMOS process \cite{ptm}, and all simulations were performed using Synpopsys HSPICE.  From this data, we estimate the energy efficiency of our design to be approximately 0.61 fJ per classification per component, where a component is either a neuron or a synapse.

\section{Quantization Approach}
\label{sec:quant}

Quantization methods for deep learning are becoming popular for accelerating training, reducing model size, and mapping neural networks to specialized hardware.  The simplest quantization methods use rounding to reduce activation and weight precision after training.  This usually results in large drops in accuracy between the full-precision and quantized models.  Other methods quantize weights, activations, and sometimes gradients during training, resulting in better performance \cite{zhou2016dorefa}.  In this work, we only quantize weights and activations.  The core idea is to use quantized values during forward propagation and full-precision gradient estimates during backward propagation.  For activations, we use a simple threshold model on the forward pass:
\begin{equation}
x=\frac{1}{2}\mathrm{sign}(s)+\frac{1}{2}
\end{equation}
where $\mathrm{sign}(\cdot)$ is 1 if the argument is non-negative and -1 otherwise.  Since the $\mathrm{sign}$ has a gradient that is zero everywhere\footnote{except when the argument approaches zero from the left, where the gradient is undefined.} it will stall the backpropagation algorithm and nothing will be learned.  To fix this, we approximate the gradient as
\begin{equation}
\frac{\partial J}{\partial s}\approx\frac{1}{1+\mathrm{exp}\left(-ks\right)}\left(1-\frac{1}{1+\mathrm{exp}\left(-ks\right)}\right),
\end{equation}
where $k$ was empirically chosen as 2.  In other words, on the backward pass, the gradient is calculated as if the activation had been a logistic sigmoid function.  Of course, we note that the threshold activation function is indeed a logistic sigmoid with a $k$ value of $+\infty$.

For weights, we use the following quantization technique:
\begin{equation}
w_{q}=2\times\frac{\mathrm{round}\left(\left(Q-1\right)\frac{\mathrm{clip}\left(w,-1,1\right)+1}{2}\right)}{Q-1}-1
\end{equation}
where $Q$ is the desired number of quantization steps, $\mathrm{round}(\cdot)$ rounds to the nearest integer and $\mathrm{clip}(w,a,b)=\mathrm{max}(a,\mathrm{min}(b,w))$, where $a\le b$.  For backpropagation, we estimate the gradient as $\partial J/\partial w\approx\partial J/\partial w_{q}$

\section{Results and Conclusions}
\label{sec:results}

\begin{figure}
\centering
\includegraphics[width=0.6\columnwidth]{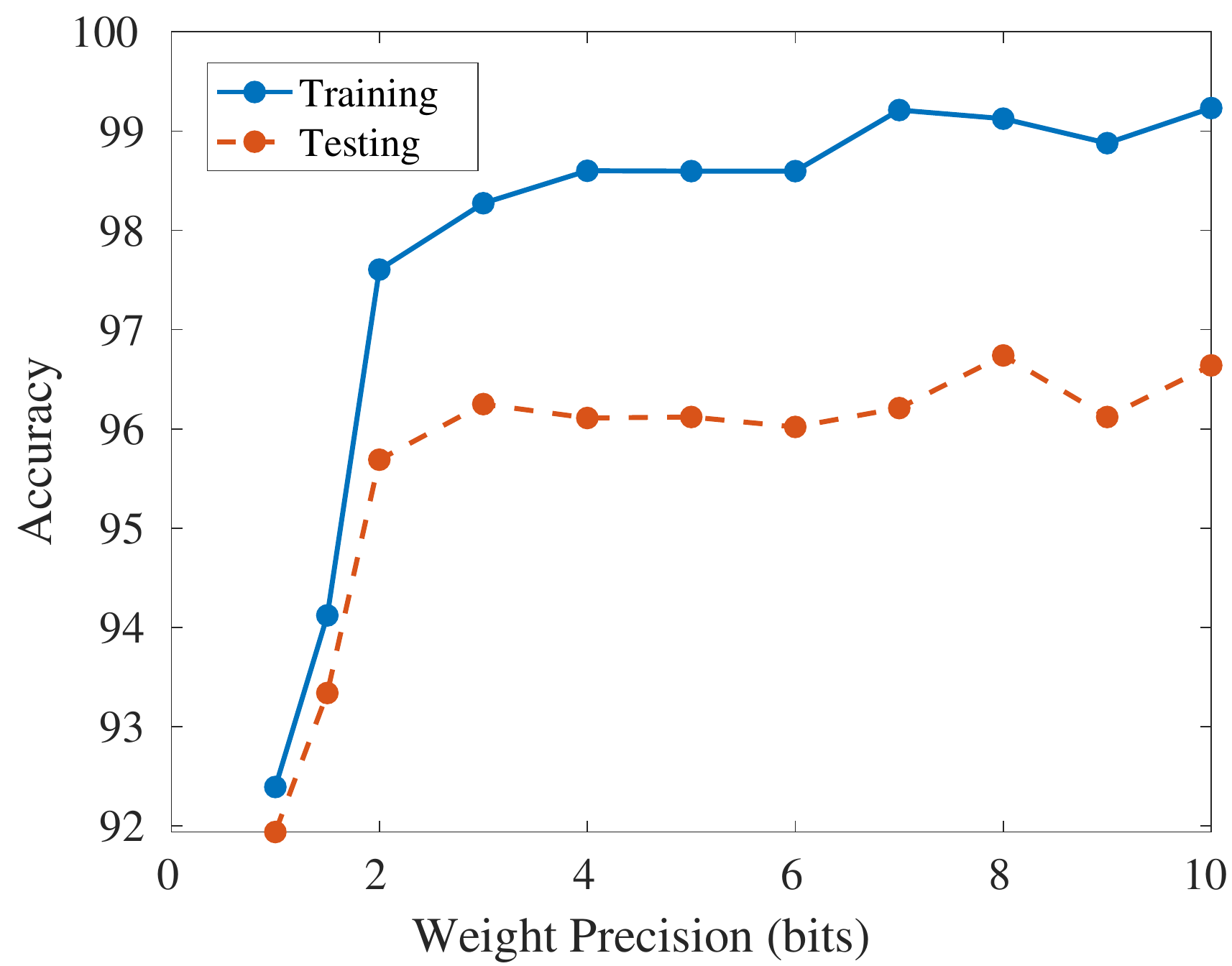}
\caption{Accuracy of the proposed design on the MNIST dataset with different weight precision.}
\label{fig:mnistaccvsprec}
\vspace{-4mm}
\end{figure}

We tested our design using the MNIST dataset of handwritten digits \cite{mnist}, which contains 60,000 training samples and 10,000 test samples of 28$\times$28 grayscale images.  Our network is parameterized with 784, 64, 64, and 10 neurons for the input, first hidden, second hidden, and output layers, respectively.  We used Tensorflow with Keras to perform all training and testing.  We have not considered any process variations in this work, so we assume that the results of Tensorflow simulations can be directly mapped to our circuit.  In the future, we plan to explore techniques for mitigating the effects of process variations using hardware-in-the-loop training.  Figure \ref{fig:mnistaccvsprec} shows the test accuracy vs. weight precision.  We observe a large increase in accuracy from 1 to 2-bit precision, which then levels off.  Note that we haven't used any regularization (dropout, etc.) in this work.  Table \ref{table:results} compares this work to other memristor-based neuromorphic systems that studied MNIST classification with low-bit weight precision.  The proposed design outperforms \cite{jiang2018pulse} by 2 orders of magnitude and is comparable to \cite{yakopcic2015memristor} in terms of energy per percent accuracy.  Power results for our design are estimated from the model presented in (\ref{eqn:power}).  Note that \cite{yakopcic2015memristor} was simulated at a 45 nm technology node, so the dynamic power would increase at 130 nm.  While these initial results are encouraging, a number of avenues for future work should be pursued to better determine the robustness of the proposed architecture, including studies on device variability and clock skew.  Also of interest for future work is the exploration of pipelined and asynchronous handshaking for coordination across layers.

\begin{table}[]
\caption{Comparison of memristor-based neuromorphic designs on MNIST classification.}
\vspace{-2mm}
\label{table:results}
\footnotesize
\begin{tabular}{cccccc}
\hline
Ref & Tech. Node & Accuracy & Power & Latency & Energy/\% Accuracy \\ \hline
\cite{jiang2018pulse}    &  130 nm  &  86\%     &   53 mW    & 80 ns &    4.93$\times 10^{-11}$ J/\%  \\
\cite{yakopcic2015memristor} & 45 nm    & 92\% &   1.79 mW    & 40 ns & 7.78$\times 10^{-13}$ J/\% \\
 This work   & 130 nm & 96\% & 0.168 mW & 400 ns & \textbf{7.01$\times \mathbf{10^{-13}}$} J/\% \\ \hline
\end{tabular}
\end{table}

\vspace{-2mm}
\bibliographystyle{ACM-Reference-Format}
\bibliography{main.bib}

\end{document}